\documentclass[11pt,a4paper]{article}
\pdfoutput=1
\usepackage{jcappub}
\usepackage{bm}
\usepackage[flushleft]{threeparttable}
\usepackage{multirow}
\usepackage{xcolor}

\def\aap{Astron.~\& Astrophys.}
\def\aj{Astron.~J.}
\def\apj{Astrophys.~J.}
\def\apjl{Astrophys.~J.~Lett.}
\def\apjs{Astrophys.~J.~Supp.}

\def\mnras{Mon.~Not.~Roy.~Astron.~Soc.}

\def\prd{Phys.~Rev.~D}
\def\prc{Phys.~Rev.~C}

\def\jcap{J.~Cosmo.~Astropart.~Phys.}
\def\pasp{Pub.~Astron.~Soc.~Pacific}

\begin{document}

\title{Reconstructing the Cosmic Expansion History with a Monotonicity Prior}
\author[a,b]{Youhua Xu}
\author[b]{Hu Zhan}
\author[a]{and Yeuk-Kwan Edna Cheung}

\affiliation[a]{School of Physics, Nanjing University,
22 Hankou Road, Nanjing, Jiangsu 210093, China}
\affiliation[b]{CAS Key Laboratory of Space Astronomy and Technology,
National Astronomical Observatories, Beijing 100012, China}

\emailAdd{yhxu@nao.cas.cn}
\emailAdd{zhanhu@nao.cas.cn}

\arxivnumber{1710.02947}

\abstract{
The cosmic expansion history, mapped by the Hubble parameter as a function
of redshift, offers the most direct probe of the dark energy equation of state.
One way to determine the Hubble parameter at different redshifts is essentially
differentiating the cosmic age or distance with respect to redshift, which may
incur large numerical errors with observational data. Taking the scenario that
the Hubble parameter increases monotonically with redshift as a reasonable prior,
we propose to enforce the monotonicity when reconstructing the Hubble parameter
at a series of redshifts. Tests with mock type Ia supernova (SN Ia) data show
that the monotonicity prior does not introduce significant biases and that
errors on the Hubble parameter are greatly reduced compared to those determined
with a flat prior at each redshift. Results from real SN Ia data are in good
agreement with those based on ages of passively evolving galaxies. Although
the Hubble parameter reconstructed from SN Ia distances does not necessarily
provide more information than the distances themselves do, it offers a convenient
way to compare with constraints from other methods. Moreover, the monotonicity
prior is expected to be helpful to other probes that measure the Hubble parameter
at multiple redshifts (e.g., baryon acoustic oscillations), and it may be generalized
to other cosmological quantities that are reasonably monotonic with redshift.
}

\maketitle

\section{Introduction} \label{label_introduction}

The Hubble parameter $H(z)$ maps the cosmic expansion history and is hence a
crucial probe of the late-time accelerated expansion of the universe revealed
by the luminosity distance-redshift relation of type Ia supernovae
(SNe Ia)~\citep{1998AJ....116.1009R,1999ApJ...517..565P}.  Dark energy and
modification to Einstein's General Relativity are the two most discussed
classes of models that try to explain the acceleration~\citep{1999PhRvL..82..896Z,
1999PhRvD..59l3504S,lrr_FR,2013CQGra..30u4003T, 2003PhRvD..68b3522S,
2001PhLB..511..265K, 2006tmgm.meet.840G,2002PhRvD..66d3507B,2010RvMP...82..451S,
2010ApJ...716..712A, 2000PhLB..485..208D}.  Dark energy and modified gravity
models generally predict different growth history of the large-scale structure
even if they produce the same cosmic expansion history.  Therefore, one can
potentially distinguish dark energy and modified gravity models by measuring
both the expansion and growth histories~\citep{2008JCAP...05..021W,
2013PhRvD..87b3520S, 2015APh....63...23H}.  Furthermore, under General Relativity,
the Hubble parameter offers the most direct probe of the dark energy equation of
state (EoS), whereas other observables are often integrals of $H^{-1}(z)$ or
solved with differential equations involving $H(z)$. In modified gravity models,
$H(z)$ is often still the most directly affected quantity by the specific
modification to gravity theory.  Measurements of the Hubble parameter are
highly complementary to other types of measurements. Future surveys such as
the Dark Energy Spectroscopic Instrument\footnote{\url{http://desi.lbl.gov/}}
(DESI), the Euclid mission\footnote{\url{http://sci.esa.int/euclid/}}, and the
Wide Field Infrared Survey Telescope\footnote{\url{http://wfirst.gsfc.nasa.gov/}}
(WFIRST) can determine $H(z)$ in different ways and cross-check with each 
other for robust tests on cosmology. 

Despite its importance, only a few practical methods are available for measuring
the Hubble parameter up to moderately high redshifts.  Spectroscopic surveys of
galaxies and quasars can determine $H(z)$ from line-of-sight baryon acoustic
oscillations (BAOs)~\citep{2003ApJ...598..720S,2003PhRvD..68f3004H,
2005ApJ...633..560E}. This will be a major undertaking of DESI, Euclid, and WFIRST.
Another method is to essentially differentiate the redshift of passively evolving
galaxies with respect to the age~\citep{2002ApJ...573...37J, 2010MNRAS.406.2569C},
which is fairly independent of the cosmological model.  However, noisy numerical
differentiation together with observational errors and astrophysical modelling errors
leads to rather large uncertainties~\citep{2005PhRvD..71j3513W, 2002ApJ...573...37J}.  
One can also differentiate SN Ia distances with respect to redshift to determine
$H(z)$. Although such an analysis does not gain more information than already
obtained from the distances, it is useful for cross-checking with other probes.
We mention in passing that the redshift drift~\citep{1962ApJ...136..319S,
1998ApJ...499L.111L} and the redshift difference between strongly lensed
images of the same object~\citep{2011ApJ...740...26Z} are sensitive to $H(z)$
as well. However, the required observations and analyses, if practical at all, are
considerably more challenging.

Instead of differentiation, one can reconstruct the Hubble parameter from the
age or distance data. Reconstruction may be parametric or
non-parametric~\citep{2006IJMPD..15.2105S,2015JCAP...09..045V}.
With the former, the quantity of interest is described by a model with a small set
of parameters, which works well if the model accurately reflects the underlying
physics. However, one could inadvertently reach a wrong conclusion because the
results are model-dependent.  In non-parametric reconstruction, the quantity of
interest is ideally represented by a complete set of orthogonal bases whose
coefficients are determined by the data. An example would be a piecewise constant
dark energy EoS in intervals of redshift or scale factor~\citep{2003PhRvL..90c1301H}.
To reduce the number of degrees of freedom,  one often truncates the basis set,
and sometimes the basis functions are only approximately orthogonal in practice.
An issue with non-parametric reconstruction is that the errors of the coefficients
obtained usually increase with the number of basis functions used and eventually
become proportional to the square root of the latter once the eigenmodes reach
stable shapes~\citep{Zhan2009}.  Therefore, care needs to be taken when comparing
results reconstructed from different datasets with different bases.

While reconstructing the Hubble parameter non-parametrically from SN Ia data with
Markov Chain Monte Carlo (MCMC) sampling, we observe that many sample points have
an $H(z)$ decreasing with redshift at least once in the whole history. This means
that the total matter-energy density of the universe increased with time at some
point in history under General Relativity, which would not happen in a flat universe
with only radiation, dust, dark matter, and the cosmological constant, i.e., the
$\Lambda$CDM model. Even if the cosmological constant is replaced by dark energy
with a constant EoS (the $w$CDM model) or an EoS changing linearly with scale
factor $a$ (the $w_a$CDM model~\citep{2001IJMPD..10..213C,2003PhRvL..90i1301L}),
it is still unusual that $H(z)$ would decrease with $z$. We therefore take a step
further proposing a monotonicity prior on $H(z)$, which requires the Hubble parameter
to be an increasing function of redshift, or, equivalently, the total matter-energy
density of the universe to always decrease with time under General Relativity.
Tests with mock SN Ia data show that the monotonicity prior improves the Hubble
parameter reconstruction without introducing significant biases. The reconstructed
$H(z)$ from existing SN Ia samples is in good agreement with that from the ages
of passively evolving galaxies.

The rest of this paper is organized as follows. Section~\ref{sec_method} outlines
the reconstruction method and discusses the monotonicity prior for the Hubble
parameter. In section~\ref{sec_application}, we carry out tests on mock SN Ia
samples and present the reconstructed $H(z)$ from existing SN Ia compilations
-- Supernova Cosmology Project (Union2.1)~\citep{2012ApJ...746...85S},
Supernova Legacy Survey three-year sample (SNLS3)~\citep{2010A&A...523A...7G,
2011ApJS..192....1C,2011ApJ...737..102S} and SDSS-II/SNLS3 Joint Light-curve
Analysis (JLA)~\citep{2013A&A...552A.124B,2014A&A...568A..22B}. An application
to WFIRST is discussed briefly in section~\ref{sec_forecast}, and we summarize
this work in section~\ref{sec_summary}.

\section{Reconstruction method}
\label{sec_method}

\subsection{From Hubble parameter to SN Ia likelihood}
\label{subsec_Hi}

We approximate the Hubble parameter as a function of redshift with 
a log-linear interpolation from its values at a series of redshifts, e.g.,
\begin{eqnarray}
\label{eqn_interp_Hi}
\hat{H}(z|\{z_i,H_i\}) =
\exp \left[ \ln H_i + \ln \left(\frac{H_{i+1}}{H_i}\right) \left(\frac{z-z_i}{z_{i+1}-z_i}\right)\right],
\end{eqnarray}
where $H_i=H(z_i)$, $z_i \leq z < z_{i+1}$, and $\{z_i\}$ is covered by 
observational data. Following the usual notation, $H_0$ is the Hubble constant.
Although eq.~\ref{eqn_interp_Hi} results in a discontinuous deceleration function,
which may not be physical, it has the advantage that a change to $H_i$ only affects
interpolated values locally between $z_{i-1}$ and $z_{i+1}$. The interpolated
deceleration and jerk functions can be made continuous with a cubic-spline.
In such a case, there would not be much difference in the reconstructed results
when the monotonicity prior is applied.

Because $H_0$ is
degenerate with SN Ia peak absolute magnitude, one can only reconstruct the cosmic
expansion function $E(z)=H(z)/H_0$ from SN Ia distances.  Hence, the actual interpolation
is performed with $\hat{E}(z|\{z_i,E_i\})$, and we take the cosmic expansion rate
$\bm{E} \equiv \{E_i\}$ as the target of reconstruction. When needed, we apply a prior
on $H_0$ to recover the constraints on $\bm{H} \equiv \{H_i\}$ from those on $\{E_i\}$.
Hereafter, we drop $\{z_i,H_i\}$ and $\{z_i,E_i\}$ in $\hat{H}(z)$ and $\hat{E}(z)$
respectively for convenience.

Under the assumption of an isotropic universe, the peak apparent magnitude $m_{\rm B}$ of a
SN Ia at redshift $z$ is given by
\begin{eqnarray} \label{eqn_mB}
m_{\rm B}(z) &=& 5\log_{10}\mathcal{D}_{\rm L}(z) + \mathcal{M}_{\rm B},
\end{eqnarray}
where $\mathcal{M}_{\rm B} \equiv M_{\rm B}+5\log_{10}\left[c/H_0\right] + 25$ is the
reduced SN Ia peak absolute magnitude, $c$ is the speed of light in vacuum, $M_{\rm B}$
is the SN Ia peak absolute magnitude in B-band, and $\mathcal{D}_{\rm L}(z)$ is the
dimensionless luminosity distance
\begin{eqnarray}
    \label{eqn_DL}
    \mathcal{D}_{\rm L}(z) = \frac{1+z}{\sqrt{\Omega_k}} 
    \sinh \left( 
    \sqrt{\Omega_k}\int_{0}^{z}\frac{d\tilde{z}}{\hat{E}(\tilde{z})}
    \right),
\end{eqnarray}
with $\Omega_k$ being the curvature density parameter at present.
For simplicity we assume a flat spacetime, i.e., $\Omega_k=0$,
in this work.

The likelihood function of the SN Ia magnitudes for a given model is
\begin{eqnarray}
\label{eqn_Ei_like}
\mathcal{L}(\bm{m}_{\rm B}^{\rm{obs}}|\bm{E}, \bm{\theta})
&=& \frac{1}{(2\pi)^{N_s/2} |\mathbf{C}|^{1/2} } 
\exp\left[-\frac{1}{2}  \Delta\bm{m}_{\rm B}^{\mathrm{T}} \cdot
  \mathbf{C}^{-1} \cdot \Delta\bm{m}_{\rm B} \right],
\end{eqnarray}
where $\Delta\bm{m}_{\rm B}$ is the difference between the observed peak apparent
magnitude $ \bm{m}_{\rm B}^{\rm{obs}}$ (after standardization) and the theoretical
value $\bm{m}_{\rm B}^{\rm{th}}$ predicted by eq.~\eqref{eqn_mB}, and $N_s$ is the
size of the SN Ia sample in use.  The observed peak apparent magnitude depends on
a set of nuisance parameters $\bm{\theta}$, which are used in the peak-light magnitude
corrections, including the light curve shape correction coefficient $\alpha$ and
color correction coefficient $\beta$ (see table~\ref{mB_correction}).  The covariance
matrix $\mathbf{C}$ has contributions from both statistical uncertainties and systematics.
We defer the discussion of $\mathbf{C}$ until section~\ref{sne_sample}.

From Bayes' theorem one has the posterior distribution for $\bm{E}$ and $\bm{\theta}$
\begin{eqnarray}
\label{eqn_Ei_post}
\mathcal{P}(\bm{E}, \bm{\theta} | \bm{m}_{\rm B}^{\rm{obs}}) \propto
\mathcal{L}(\bm{m}_{\rm B}^{\rm{obs}}|\bm{E}, \bm{\theta})
\, \pi(\bm{E}) \, \pi(\bm{\theta}),
\end{eqnarray}
where $\pi(\bm{E})$ and $\pi(\bm{\theta})$ are the priors, and $\bm{E}$ and
$\bm{\theta}$ are assumed to be uncorrelated. The reduced peak absolute magnitude
$\mathcal{M}_{\mathrm{B}}$ is absorbed into $\bm{\theta}$ for notational simplicity.
The normalization factor or the Bayesian evidence of the posterior distribution is
useful for model selection~\citep{1538-4357-638-2-L51,2007MNRAS.378.72T}. For our
purpose, we treat it as an irrelevant constant.

\subsection{Monotonicity prior}
\label{subsection_prior}

When a flat prior (also known as non-informative prior) is applied to the parameters,
one gets many unusual sample points that have an expansion rate decreasing with
redshift at some point. Expansion histories represented by these sample points are
unlikely to have happened in the concordant $\Lambda$CDM cosmological model or its
extensions such as the $w$CDM and $w_a$CDM models. To remove these rather unphysical
sample points, we propose to apply a monotonicity prior on the Hubble parameter,
which requires $H(z)$ to be a function monotonically increasing with redshift.
It is equivalent to the Null Energy Condition in a flat universe \citep{2008PhRvD..77h3518L,2015arXiv150200811B}.
Since $H(z)=H_0\, E(z)$, the prior on the Hubble parameter may be written as
a product of two uncorrelated parts: $\pi(\bm{H}) = \pi(H_0) \, \pi(\bm{E})$. 
Thus the monotonicity prior reads
\begin{eqnarray} \label{eq_MP_E}
\pi(\bm{E}) =
\left \{\begin{array}{lll}
		1, &~ & \mbox{$E_i \le E_{i+1}$ for all $i$},\\
		0, &~ & \mbox{otherwise}.
	    \end{array}
\right.
\end{eqnarray}
To obtain constraints on the Hubble parameter, one maps the reconstructed cosmic
expansion rate $\bm{E}$ into Hubble rate $\bm{H}$ by multiplying every sample point
in the $\bm{E}$ chains with a randomly sampled $H_0$ from a prior distribution
$\pi(H_0)$, e.g., ref.~\citep{0004-637X-826-1-56}. In appendix~\ref{prior_eta} we
provide an equivalent form of the monotonicity prior that is more efficient for
MCMC sampling.

In the $\Lambda$CDM model, the expansion function is
$E(z;\Omega_{\rm m}) = \sqrt{\Omega_{\rm m}(1+z)^3 + (1-\Omega_{\rm m})}$, where
$\Omega_{\rm m}$ is the matter density parameter defined as the ratio of today's
matter density to the critical density $\rho_{\rm{crit}} \equiv 3H^2_0/8\pi G$ with
$G$ being the Newton's gravitational constant.  The derivative of the expansion
function with respect to redshift is
$E'(z;\Omega_{\rm m})=3\Omega_{\rm m}(1+z)^2/2E(z;\Omega_{\rm m})$,
which automatically satisfies the monotonicity prior.

In the $w$CDM model, the expansion function is
\begin{eqnarray}
E(z;\Omega_{\rm m}, w)= \sqrt{\Omega_{\rm m}(1+z)^3 + (1-\Omega_{\rm m})(1+z)^{3(1+w)}},
\end{eqnarray}
and its derivative with respect to redshift is
\begin{eqnarray} \label{eqn_dE_wcdm}
E'(z;\Omega_{\rm m}, w) &=& \frac{3(1+z)^2}{2} 
	\frac{\Omega_{\rm m} + (1+w)(1-\Omega_{\rm m})(1+z)^{3w} }{E(z;\Omega_{\rm m}, w)}.
\end{eqnarray}
Because $\Omega_{\rm m}$ lies between zero and unity, $E'(z;\Omega_{\rm m},w) \ge 0$
will always be satisfied if $w$ is no less than $-1$.  When $w<-1$, $E'(z;\Omega_{\rm m}, w)$
remains non-negative if
\begin{eqnarray} \label{eqn_wcdm}
\frac{\Omega_{\rm m}} {1-\Omega_{\rm m}} &\ge& -(1+w)(1+z)^{3w}.
\end{eqnarray}
Since the term $(1+z)^{3w}$ decreases with increasing $z$, the right side of 
eq.~\eqref{eqn_wcdm} reaches its maximum at $z=0$.  Thus the inequality holds
at any redshift as long as $\Omega_{\rm m} (1-\Omega_{\rm m})^{-1} \ge -(1+w)$.

\begin{figure}[t!]
\centering
\includegraphics[width=0.65\textwidth]{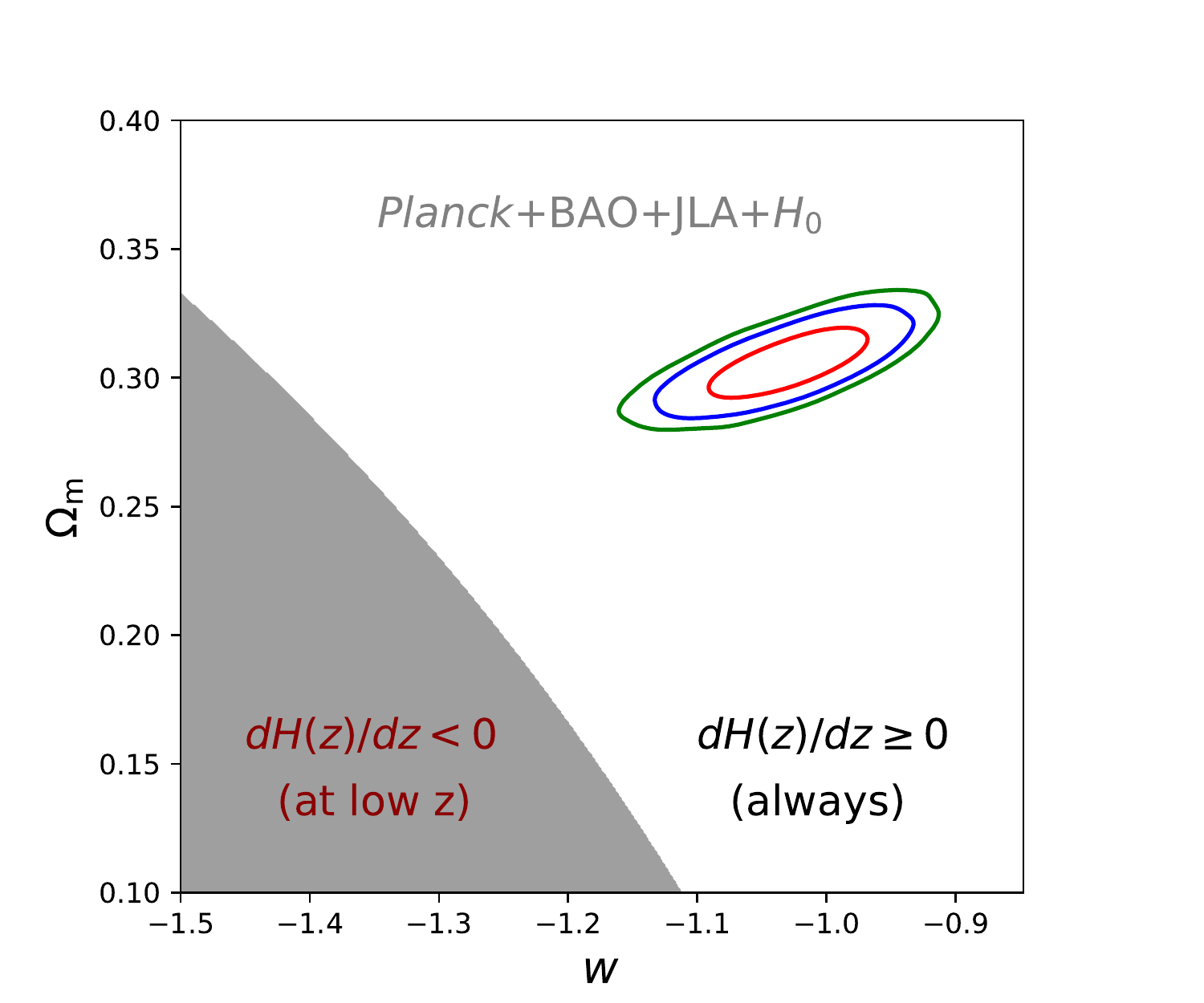}
\caption{Consistency check of the  monotonicity prior in the $w$CDM
  cosmology. The contours are joint constraints of {\it Planck}, BAO,
  JLA, and $H_0$ \citep{Planck2015_cos_par} and correspond to $68\%$,
  $95\%$, and $99\%$ confidence levels from inside out. The shaded area
  shows the sub-space rejected by the monotonicity prior. }
\label{fig_prior_wcdm}
\end{figure}

In figure~\ref{fig_prior_wcdm}, we plot the marginalized constraints on $w$ and
$\Omega_{\rm m}$ (the contours) from the combination of {\it Planck}, BAO,
JLA, and $H_0$ \citep{Planck2015_cos_par} and show the $w$--$\Omega_{\rm m}$
sub-space that is rejected by the monotonicity prior (shaded area).
As one sees from the figure, the combined observations rule out the possibility
of an $E'(z;\Omega_{\rm m},w) < 0$ expansion history with a wide margin.  Therefore,
the monotonicity prior is reasonable within the $w$CDM model.

The EoS evolves with redshift in the $w_a$CDM model and is parameterized by
$w(z)=w_0+w_a \frac{z}{1+z}$~\citep{2001IJMPD..10..213C,2003PhRvL..90i1301L}.
The expansion function is
\begin{eqnarray}
E(z;\Omega_{\rm m},w_0, w_a) = \sqrt{\Omega_{\rm m}(1+z)^3 + (1-\Omega_{\rm m})f_{X}(z)},
\end{eqnarray}
where
\begin{eqnarray}
f_{X}(z) \equiv \exp \left[ 3\int_0^z \frac{1+w(z')}{1+z'}dz'\right].
\end{eqnarray}
The derivative of the expansion function with respect to redshift is now given by
\begin{eqnarray}
E'(z;\Omega_{\rm m},w_0, w_a) &=& \frac{3}{2} 
\frac{\Omega_{\rm m}(1+z)^3 + (1-\Omega_{\rm m})[1+w(z)] f_X(z)}{(1+z)E(z;\Omega_{\rm m},w_0, w_a)}.
\end{eqnarray}

\begin{figure}[t!]
\centering
\includegraphics[width=\textwidth]{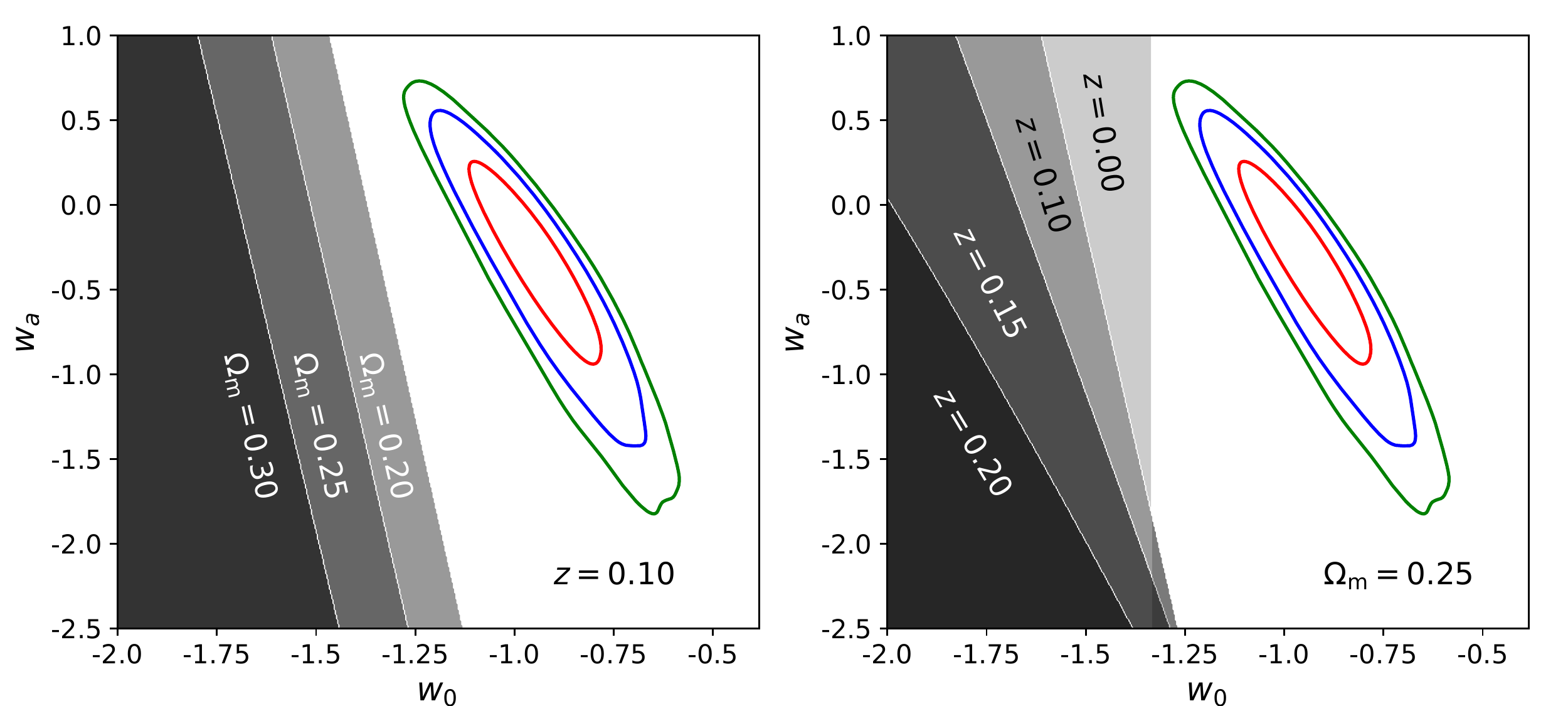}
\caption{Same as figure~\ref{fig_prior_wcdm} but for the $w_a$CDM cosmology.
Shaded areas in the left panel represent the sub-space rejected by the monotonicity prior,
at $z=0.1$ for different matter densities; those in the right panel represent the rejected
sub-space at different redshifts for fixed matter density $\Omega_{\rm m}=0.25$.}
\label{fig_prior_wacdm}
\end{figure}

Figure~\ref{fig_prior_wacdm} shows the marginalized constraints on $w_0$
and $w_a$ (contours) from the same set of data used in figure~\ref{fig_prior_wcdm}.
Shaded areas in the left panel are the sub-space rejected by the monotonicity prior for
different matter densities at $z=0.1$, and those in the right panel are the sub-space
rejected by the monotonicity prior at different redshifts for $\Omega_{\rm m}=0.25$.
As the matter density increases, the sub-space allowed by the monotonicity prior becomes
larger and larger.  From the right panel one finds that the sub-space allowed by the
monotonicity prior increases with redshift for a fixed matter density.  In all cases shown
in figure~\ref{fig_prior_wacdm}, the monotonicity prior is consistent with existing
observations under the $w_a$CDM model.

Although we cannot enumerate all possible models, it is reasonable to take as 
a prior that the Hubble parameter increases monotonically with redshift.

\subsection{Affine invariant MCMC ensemble sampling }
\label{mcmc_algorithm}

We employ the affine invariant MCMC ensemble sampler proposed by Goodman $\&$
Weare~\citep{goodman_weare} to explore the parameter space.  This sampler's
algorithm has been demonstrated to be much more efficient than the traditional
ones, e.g., the Metropolis-Hastings algorithm~\citep{2012ApJ...745..198H,
2013PASP..125..306F,Akeret:2012ky}.  During the course of sampling, only one
single position is updated in the Metropolis-Hastings algorithm, whereas an
ensemble of $K$ walkers $W=\{X_k\}$ are evolved simultaneously in the affine
invariant MCMC.  The proposal distribution for the $k$-th walker is determined
by the current positions of the other $K-1$ walkers in the complementary ensemble
$W_{\left[k\right]}= \{X_j,\forall j\neq k \}$, where the positions refer to vectors in
the $N_{\rm p}$-dimensional parameter space.  To update the position of a walker $X_k$,
another walker $X_j$ is randomly drawn from the remaining walkers $W_{\left[k\right]}$ and
a new position is proposed to be
\begin{eqnarray}
	\label{emcee_proposal}
	X_k \rightarrow Y=X_j + R (X_k-X_j),
\end{eqnarray}
where $R$ is a random variable drawn from a distribution $g(r)$.  The
distribution $g(r)$ is required to have the following property
\begin{eqnarray}
	g(r^{-1}) = r ~g(r),
\end{eqnarray}
so that the proposal of eq.~\eqref{emcee_proposal} is symmetric.  An example
for $g(r)$ is provided in~\citep{goodman_weare, 2013PASP..125..306F}
\begin{eqnarray}
	\label{eqn_g}
	g(r) \propto
	\left \{ \begin{array}{rr}
		\displaystyle \frac{1}{\sqrt{r}}, &\mbox{if $r$ $\in \left[s^{-1}, s\right]$},\\
		0,  & \mbox{otherwise},
		\end{array}
	\right.
\end{eqnarray}
where $s$ is an adjustable scale parameter controlling the jumping step sizes
of the walkers.  The chains satisfies the condition of detailed balance if the
proposed position $Y$ is accepted with the following probability
\begin{eqnarray}
	\label{eqn_q}
	q = \min\left( 1, R^{N_\mathrm{p}-1}\frac{p(Y)}{p(X_k)} \right),
\end{eqnarray}
where $N_\mathrm{p}$ is the dimension of the parameter space.  This procedure is then
repeated for each of the rest walkers in the ensemble.

Two features of the affine invariant MCMC are worth attention.  The first is
that it is suitable for cases where parameters are strongly correlated with
each other.  Such correlations often cause high rejections of the proposed
steps and thus lower the sampling efficiency of conventional algorithms.
The affine invariant MCMC works well in these cases because of its
affine-invariant property, and the sampling efficiency or acceptance ratio
is controlled by a single parameter, the scale parameter $s$ in eq.~\eqref{eqn_g}.
The second is that it can be naturally parallelized, which is crucial for
computationally intensive problems, as shown by the pseudo-code in
ref.~\citep{2013PASP..125..306F}.

We implement a C++ version\footnote{\url{https://xyh-cosmo.github.io/imcmc/}.}
of the sampler to reconstruct the Hubble parameter via MCMC.  The chains produced
by our sampler share the same format as that of CosmoMC\footnote{\url{http://cosmologist.info/cosmomc/}.},
and they can be readily processed with the python package GetDist\footnote{\url{https://github.com/cmbant/getdist}.}
to get the best-fit values of the parameters and their corresponding confidence intervals.

\section{Expansion history from SN Ia distances}
\label{sec_application}

In this section, we test the monotonicity prior using simulated SN Ia mock
samples and reconstruct the cosmic expansion history from three existing SN
Ia samples.  We then apply the method to the expected WFIRST SN Ia sample and
compare the resulting constraints on the cosmic expansion rate $E(z)$ with
those from DESI BAO.

\subsection{SN Ia samples}
\label{sne_sample}

The SN Ia samples to be used for reconstruction include Union2.1~\citep{2012ApJ...746...85S},
SNLS3~\citep{2011ApJS..192....1C, 2010A&A...523A...7G} and JLA~\citep{2014A&A...568A..22B,2013A&A...552A.124B}.
These compilations use the SALT2 model~\citep{2005A&A...443..781G, 2007A&A...466...11G}
to fit the SNe Ia light-curves, from which the peak rest-frame B-band apparent
magnitude $m_{\rm B} $ at the epoch of maximum light, the shape of the
light-curve $S$ and the optical B-V color $C$ are measured.  These quantities
are then used to standardize the SN Ia peak luminosity.  The measured distance
modulus after correction can be written as a linear combination of the above
measured quantities
\begin{eqnarray}
\mu_B = m_{\rm B} - M_{\rm B} + \alpha\, S - \beta\, C,
\label{eq_sn_mu}
\end{eqnarray}
where the B-band absolute magnitude $M_{\rm B}$ and the coefficients $\alpha$
and $\beta$ are fitted together with cosmological parameters. SNLS3 also uses
SiFTO model~\citep{2011ascl.soft10023C} to fit the SNe Ia light-curves, but
the shape parameter in eq.~\eqref{eq_sn_mu} is replaced by $s-1$. In
table~\ref{mB_correction} we summarize the specific light-curve parameterizations
adopted by these compilations: Union2.1 and JLA used exactly the same form as in
eq.~\eqref{eq_sn_mu}, whereas SNLS3 used the SiFTO parameterization; a detailed
comparison of SLAT2 and SiFTO can be found in ref.~\citep{2010A&A...523A...7G},
where transformations that relate the parameterizations of SiFTO and SLAT2 are
also given.  

\begin{table}[t!] \renewcommand{\arraystretch}{1.5}
\centering
\begin{threeparttable}
\caption{SN Ia Peak-light magnitude corrections.}
\label{mB_correction}
\begin{tabular}{c|rl|c}
\hline \hline
SN Ia Compilation & &
Magnitude Corrections\tnote{a}
& Nuisance Parameters \\
\hline
Union2.1 & 
$\mu_{\rm B}=$ & $m_{\rm B}+\alpha\, x_1-\beta\, c+\delta\, P-M_{\rm B}$ &
$\alpha, \beta, M_{\rm B}, \delta$ \\
\hline
SNLS3      & $m_{\rm B}=$ &
$5\log_{10}\mathcal{D}_{\rm L}-\alpha (s-1)+\beta\,C+\mathcal{M}_{\rm B}$ &
$\alpha,\beta,\mathcal{M}_{\rm B}$ (or $\mathcal{M}^1_{\rm B},\mathcal{M}^2_{\rm B}$) \\
\hline
JLA      &
$\mu_{\rm B}=$ &
$m_{\rm B}-(M_{\rm B}-\alpha\, X_1 + \beta\, C)$ &
$\alpha,\beta,M_{\rm B}$ (or $M_{\rm B}^1, \Delta_{\rm M}$) \\
\hline \hline
\end{tabular}
\begin{tablenotes}
\item[a] These expressions have been simplified for clarity.
\end{tablenotes}
\end{threeparttable}
\end{table}

The SN Ia luminosity slightly correlates with the host galaxy
mass~\citep{2010MNRAS.406..782S}.  This correlation is accounted for by
introducing new parameters to the peak-light magnitude corrections, as shown
in table~\ref{mB_correction}.  Union2.1 models this correlation with the term
$\delta\, P$, where $\delta$ is the host-mass-correction coefficient and $P$
is the probability for a SN Ia to have a host galaxy less massive than
$10^{10}m_\odot$.  SNLS3 and JLA divide their SNe Ia into two sub-samples,
one for SNe Ia with host galaxies less massive than $10^{10} m_{\odot}$ and
the other for those with more massive hosts.  SNLS3 assigns two reduced peak
absolute magnitudes, $\mathcal{M}_{\rm B}^1$ and $\mathcal{M}_{\rm B}^2$, to
its two sub-samples, respectively; JLA uses a single peak absolute magnitude
parameter $M_{\rm B}^1$ for all SNe Ia, and adds an offset $\Delta_{\rm M}$
to $M_{\rm B}^1$ for those SNe Ia in more massive host galaxies.

There are many systematics that affect the constraints on cosmological parameters.  
The zero-point uncertainties and the absolute magnitude calibration are the dominant
effects.  Others include Malmquist bias, galactic and intergalactic extinction, 
rest-frame U-band calibration, non-SN Ia contamination as well as gravitational
lensing~\citep{2010ApJ...716..712A,2012ApJ...746...85S,2011ApJS..192....1C,
2010A&A...523A...7G, 2014A&A...568A..22B, 2013A&A...552A.124B}.  All the three
compilations propagate these systematics into the total covariance matrices.
SNLS3 and JLA further take into account the systematics induced by peculiar velocities
and possible evolution of SN Ia.

The total covariance matrix can be decomposed into a summation of three
terms following the convention in ref.~\citep{2010A&A...523A...7G}:
\begin{eqnarray}
\mathbf{C} = \mathbf{D}_{\rm{stat}}+\mathbf{C}_{\rm{stat}}+\mathbf{C}_{\rm{sys}}.
\end{eqnarray}
The off-diagonal parts $\mathbf{C}_{\rm{stat}}$ and $\mathbf{C}_{\rm{sys}}$ are
the statistical and systematical correlations between different SNe Ia, and they
can be computed using the standard techniques described in ref.~\citep{2005PhRvC..72e5502A}.  
The diagonal matrix $\mathbf{D}_{\rm{stat}}$ represents statistical
uncertainties of the SNe Ia
\begin{eqnarray}
\mathbf{D}_{{\rm{stat}},ii}&=&
\sigma^2_{m_{\rm B},i}
+\alpha^2\sigma_{s,i}^2
+\beta^2\sigma_{C,i}^2
+\sigma^2_{\rm{int}}
+\left(\frac{5(1+z_i)}{z_i(1+z_i/2)\log{10}}\right)^2\sigma^2_{z,i}\nonumber\\
& + & \sigma^2_{\rm{lens}}+\sigma^2_{\rm{host}}+{\mathbf{C}}_{m_B\,s\,C,i},
\label{cov_Dii}
\end{eqnarray}
where $\sigma^2_{m_{\rm B},i}$, $\sigma_{s,i}^2$
and $\sigma_{C,i}^2$ are the variances of the measured light-curve
parameters, $\sigma^2_{\rm{int}}$ accounts for the intrinsic scatter of SN Ia, 
the term proportional to $\sigma^2_{z,i}$ is the magnitude variance
associated with redshift uncertainty
(with $z_i$ being the redshift of the $i$th SN Ia in the CMB rest frame), 
$\sigma^2_{\rm{lens}}$ is caused by gravitational lensing,  
$\sigma^2_{\rm{host}}$ comes
from the correlation of SN Ia luminosity to the stellar mass of its host galaxy, the last term
${\mathbf{C}}_{m_B\,s\,C,i}$ represents three covariances between the light-curve
parameters as well as the apparent magnitude for each SN Ia and is a function of $\alpha$
and $\beta$.   The (total) statistical uncertainties depend on the light-curve parameters
$\alpha$ and $\beta$, which can not be analytically marginalized and are fitted together
with the cosmological parameters (the strategy used in SNLS3 and JLA); Union2.1 provides
total covariance matrices with and without systematics, in which $\alpha$ and $\beta$
(as well as $\delta$) assume their best-fit values.

The intrinsic scatter term $\sigma_{\rm{int}}$ is sourced from two parts.  One is the
true dispersion in the peak absolute luminosity reflecting the imperfection of these
standardized candles, and the other includes systematics that are currently unknown.
Unfortunately, these two parts can not be distinguished from each other.  To obtain an
estimate of $\sigma_{\rm{int}}$, one usual method is to fit the data to a fiducial model
(the flat $w$CDM model is often adopted) and adjust $\sigma^2_{\rm{int}}$ until the
best-fit $\chi^2$ to be one per degree of freedom.  However, this procedure precludes
the statistical tests of the adequacy of the cosmological model to describe the
data~\citep{2011ApJS..192....1C,2014A&A...568A..22B}.  The unknown  systematics
are generally sample-dependent, and SNLS3 and JLA compilations have estimated 
$\sigma_{\rm{int}}$ for each of their sub-samples (see table~\ref{table_intdisp}).
Although Union2.1 has estimated $\sigma_{\rm{int}}$ for its 19 sub-samples,
only the median value $\sigma_{\rm{int}}^{\rm median}=0.15$ is used
to construct the covariance matrix~\citep{2012ApJ...746...85S}.

\begin{table}[t!]\renewcommand{\arraystretch}{1.5}
\centering
\caption{Intrinsic scatter $\sigma_{\rm{int}}$ of the sub-samples in
SNLS3~\citep{2011ApJS..192....1C} and JLA~\citep{2014A&A...568A..22B}.}
\label{table_intdisp}
\vspace{0.2cm}
\begin{threeparttable}
\begin{tabular}{c|c|c}
\hline \hline
Sub-sample & SNLS3 & JLA  \\
\hline
Low-z & 0.113 & 0.134 \\
\hline
SDSS & 0.099 & 0.108 \\
\hline
SNLS & 0.068 & 0.080 \\
\hline
HST & 0.082 & 0.100 \\
\hline \hline
\end{tabular}
\end{threeparttable}
\end{table}

\subsection{Tests with mock samples}
\label{test_method}

We construct two mock SN Ia samples to test the reconstruction method.  The
first one is generated from the real JLA SN Ia sample, sharing exactly
the same redshifts as those of JLA~\citep{2014A&A...568A..22B,2013A&A...552A.124B}
(hereafter JLA-mock).  To simulate random magnitude errors, a full covariance
matrix for all the 740 SNe Ia is constructed from the covariance matrices
provided by the JLA compilation, with the light-curve nuisance parameters
$\alpha$ and $\beta$ fixed to 0.13 and 3.1, respectively.  The errors are
then randomly drawn from the multivariate Gaussian distribution (with zero
means) specified by the total covariance matrix, which is also used in the
JLA-mock likelihood calculations.  

The second one assumes the redshift distribution expected from
WFIRST~\citep{2015arXiv150303757S}, which will be able to measure over
2700 SNe Ia in the redshift range $0.1 \le z \le 1.7$ (hereafter WFIRST-mock).
The magnitude errors are randomly drawn from the WFIRST Gaussian error
model~\citep{2015arXiv150303757S}
\begin{eqnarray}
    \label{eqn_sigma_mu}
    \sigma_{\mu}^2(z) = \sigma^2_{\rm{int}} + \sigma^2_{\rm{meas}}
    + \sigma^2_{\rm{lens}} + \sigma^2_{\rm{sys}},
\end{eqnarray}
where $\sigma_{\rm{int}} = 0.09$ is the intrinsic spread of the SN Ia peak
absolute magnitude (after correcting for the light-curve shape and spectral
properties), $\sigma_{\rm{meas}}= 0.08$ is the photometry error,
$\sigma_{\rm{lens}}= 0.07 z$ accounts for lensing magnification, and
$\sigma_{\rm{sys}}=0.02(1+z)/1.8$ is the assumed systematics.  The magnitude
errors of different SNe Ia in this mock are assumed to be uncorrelated for simplicity.

The two mock SN Ia samples are generated assuming a flat $\Lambda$CDM cosmology
with the following fiducial parameters: $H_0=70.0~\rm{km~s^{-1}~Mpc^{-1}}$,
$\Omega_{\rm m}=0.3$, and $\Omega_\Lambda=0.7$.  The fiducial value of the SN Ia
peak absolute magnitude is assumed to be $M_{\rm B}=-19.3$.  In the following tests,
the reduced peak absolute magnitude is fitted together with the cosmic expansion
rate $\{E_i\}$ and then marginalized. 

\begin{figure}[t!]
\centering
\includegraphics[width=0.8\textwidth]{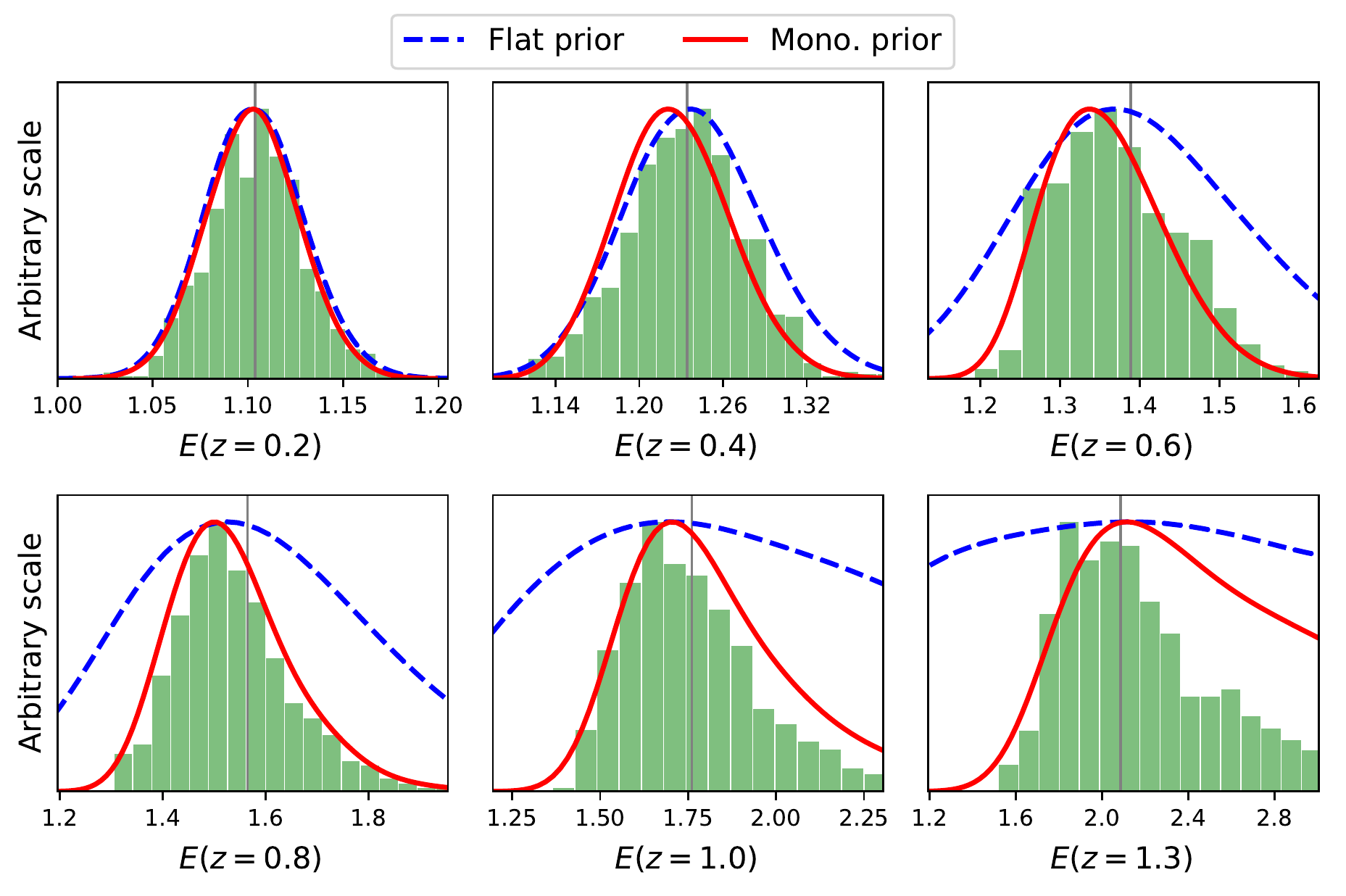}
\caption{Marginalized posterior distributions of the cosmic expansion rate
$\{E_i\}$ at different redshifts from the JLA-mock (curves) and distributions
of the maximum-likelihood $\{E_i\}$ values from 1000 random realizations 
of mock JLA data with the monotonicity prior (histograms). The solid and
dashed curves are the results reconstructed with and without the monotonicity
prior, respectively. The results are all normalized to the same peak height.  
 The vertical lines indicate the input values.}
\label{fig_mock_jla}
\end{figure}

The marginalized posterior distributions of the cosmic expansion rate $\{E_i\}$
reconstructed from the JLA-mock is shown in figure~\ref{fig_mock_jla}.  The
reconstructed results recover the input values (vertical lines) fairly well, and the
reconstruction uncertainties are reduced considerably by the monotonicity prior
(solid curves) at $z\gtrsim0.6$. Since the distance modulus $\mu(z)$ involves
an integration over $\hat{E}^{-1}(z)$ from $0$ to $z$, the cosmic expansion
rate at the first few interpolation nodes are constrained by not only the low
redshift SNe Ia but also those at higher redshifts.  Therefore the cosmic
expansion rate at low redshifts can still be well constrained even without
the monotonicity prior. 

Although the results of the particular mock data do not appear to be biased,
the ensemble behavior of the estimator can differ (e.g.,~\citep{2013ApJ...777...75S}).
Biases may be present in the distributions of values estimated from an ensemble
of realizations. One can determine an optimal balance between the ensemble bias
and the ensemble variance with a large number of random mocks (e.g.,~\citep{2015JCAP...09..045V}).
For the purpose of this work, we simply draw 1000 realizations of mock JLA data
and examine the probability distributions of the maximum-likelihood values of the
expansion rate obtained with the monotonicity prior. The resulting histograms in 
figure~\ref{fig_mock_jla} are skewed in some cases, but the biases are small
compared to the posterior distributions from the single JLA-mock, which
characterize parameter uncertainties in the usual MCMC analysis. 

\begin{figure}[t!]
\centering
\includegraphics[width=\textwidth]{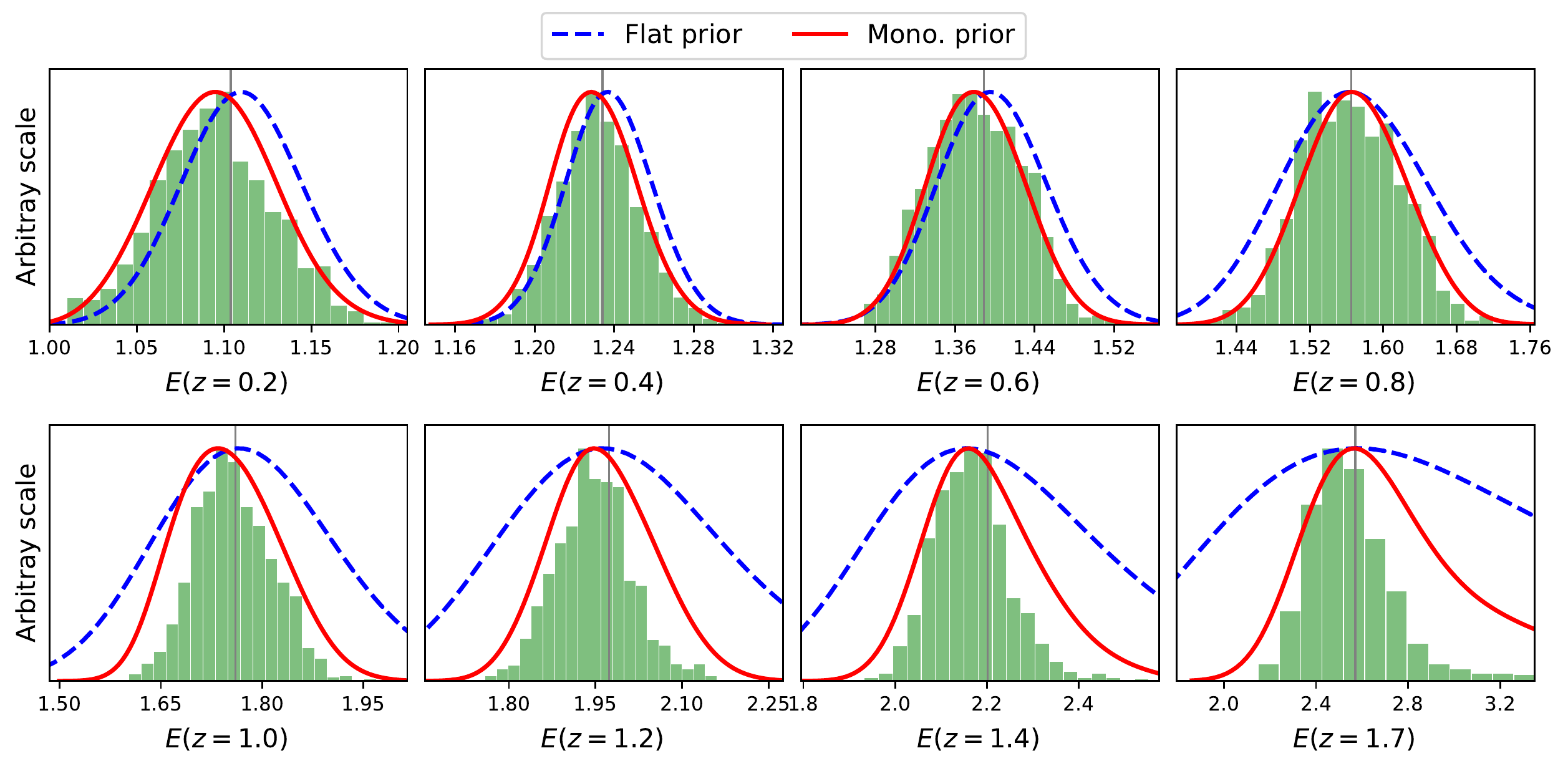}
\caption{Same as figure~\ref{fig_mock_jla} but for the WFIRST-mock.  Since the
WFIRST-mock extends to redshift $z=1.7$, two more interpolation nodes are
added.}
\label{fig_mock_wfirst}
\end{figure}

Figure~\ref{fig_mock_wfirst} shows the expansion rate $\{E_i\}$ reconstructed
from the WFIRST-mock.  One sees the same improvements as those in
figure~\ref{fig_mock_jla}.  Since the WFIRST-mock contains about 300 SNe 
Ia in the range $1.4 < z \le 1.7$, a lot more than those in the JLA-mock, 
the expansion rate at the last interpolation node, $E(z=1.7)$, can still be
constrained fairly well. At low redshift, however, the JLA-mock has many more
SNe Ia than the WFIRST-mock and hence provides a tighter constraint at $z=0.2$.
Similar to the case of JLA, the distributions of the estimated $\{E_i\}$ values from
1000 realizations of mock WFIRST data are not biased appreciably.

\begin{figure}[t!]
\centering
\includegraphics[width=0.8\textwidth]{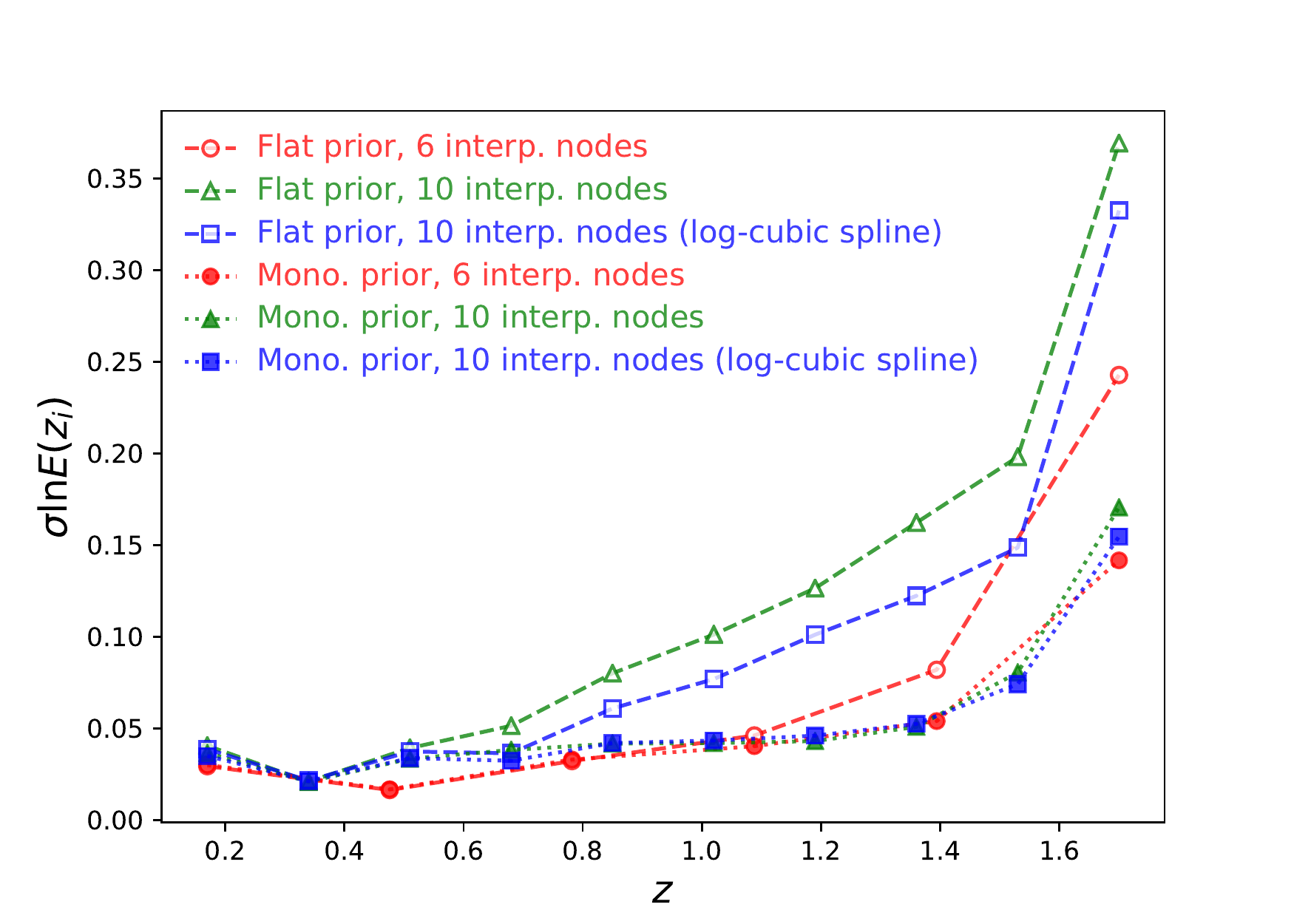}
\caption{Fractional errors of the cosmic expansion rate $\{E_i\}$ reconstructed
from the WFIRST-mock: circles for 6-node log-linear interpolation, triangles 
for 10-node log-linear interpolation, and squares for 10-node log-cubic 
spline interpolation. Open symbols represent results with the flat prior,
and solid symbols represent those with the monotonicity prior.}
\label{fig_bin_num_effect}
\end{figure}

Figure \ref{fig_bin_num_effect} compares the fractional errors on the cosmic
expansion rate reconstructed from the WFIRST-mock with 6-node (circles) and
10-node (triangles) log-linear interpolations as defined by eq.~\ref{eqn_interp_Hi}.
In addition, we also include the results of 10-node log-cubic spline interpolation
(squares) to check the effect of the interpolation scheme on the reconstruction.

When the flat prior is applied (open symbols), the fractional errors at redshifts
$z\gtrsim 0.7$ increase quickly with the number of interpolation nodes. The
interpolated deceleration and jerk functions are continuous with the log-cubic
spline interpolation, which is effectively an additional constraint on the expansion
rate. Therefore, the 10-node log-cubic spline interpolation achieves smaller
uncertainties of $\{E_i\}$ than the 10-node log-linear interpolation does.
When the monotonicity prior is enforced (solid symbols), the fractional errors
are significantly reduced and remain less than $5\%$ up to redshift $z\simeq 1.4$.
It is somewhat surprising that the errors are fairly independent of the number of
interpolation nodes and interpolation schemes. Because the added interpolation
nodes limit the allowable variation of the Hubble parameter at each redshift, the
strength of the monotonicity prior actually increases with the number of interpolation
nodes assigned to the same redshift range, largely canceling the effect of
increasing number of interpolation nodes. 

\subsection{Expansion history from existing SN Ia samples}
\label{rec_exp}

\begin{table}[t!]\renewcommand{\arraystretch}{1.5}
\centering
\caption{Cosmic expansion rate and Hubble parameter (${\rm km~s^{-1} Mpc^{-1}}$)
reconstructed from Union2.1, SNLS3 and JLA.}
\vspace{0.2cm}
\begin{threeparttable}
\begin{tabular}{c|c|c|c|c|c|c}
\hline \hline
\multirow{2}{*}{$z_i$} & \multicolumn{2}{c|}{Union2.1$^{a}$} &
\multicolumn{2}{c|}{SNLS3$^{a}$} & \multicolumn{2}{c}{JLA$^{a}$}\\
\cline{2-7}
& $E(z_i)$ & $H(z_i)$ & $E(z_i)$ & $H(z_i)$&$E(z_i)$ & $H(z_i)$\\
\hline
0.2 &
$1.09_{-0.03}^{+0.04}$ & $79.5_{-2.9}^{+3.3}$ & 
$1.08_{-0.03}^{+0.03}$ & $79.0_{-2.5}^{+2.7}$ & 
$1.11_{-0.03}^{+0.02}$ & $81.5_{-2.6}^{+2.5}$\\
\hline
0.4 &
$1.21_{-0.06}^{+0.06}$ & $88.8_{-4.3}^{+4.9}$ &
$1.12_{-0.03}^{+0.04}$ & $82.2_{-3.3}^{+3.5}$ &
$1.22_{-0.04}^{+0.04}$ & $89.1_{-3.5}^{+3.7}$\\
\hline
0.6 &
$1.35_{-0.07}^{+0.09}$ & $99.1_{-5.7}^{+6.5}$ &
$1.19_{-0.06}^{+0.07}$ & $87.1_{-4.8}^{+5.9}$ &
$1.34_{-0.06}^{+0.08}$ & $98.1_{-5.0}^{+6.2}$\\
\hline
0.8 &
$1.48_{-0.10}^{+0.12}$ & $109_{-7.0}^{+9.0}$ &
$1.35_{-0.10}^{+0.13}$ & $99.0_{-8.0}^{+10}$ &
$1.43_{-0.08}^{+0.09}$ & $105_{-6.0}^{+7.0}$\\
\hline
1.0 &
$1.66_{-0.15}^{+0.23}$ & $122_{-12}^{+17}$ &
$1.56_{-0.17}^{+0.27}$ & $115_{-13}^{+20}$ &
$1.57_{-0.13}^{+0.21}$ & $115_{-10}^{+16}$\\
\hline \hline
\end{tabular}
\begin{tablenotes}
\item[a]{The results are obtained with the monotonicity prior. The
  quoted values correspond to the peak positions of the marginalized
  one-dimensional distributions and $68\%$ confidence levels.}
\end{tablenotes}
\end{threeparttable}
\label{table_USJ_Ei}
\end{table}

\begin{figure}[t!]
\centering
\includegraphics[width=\textwidth]{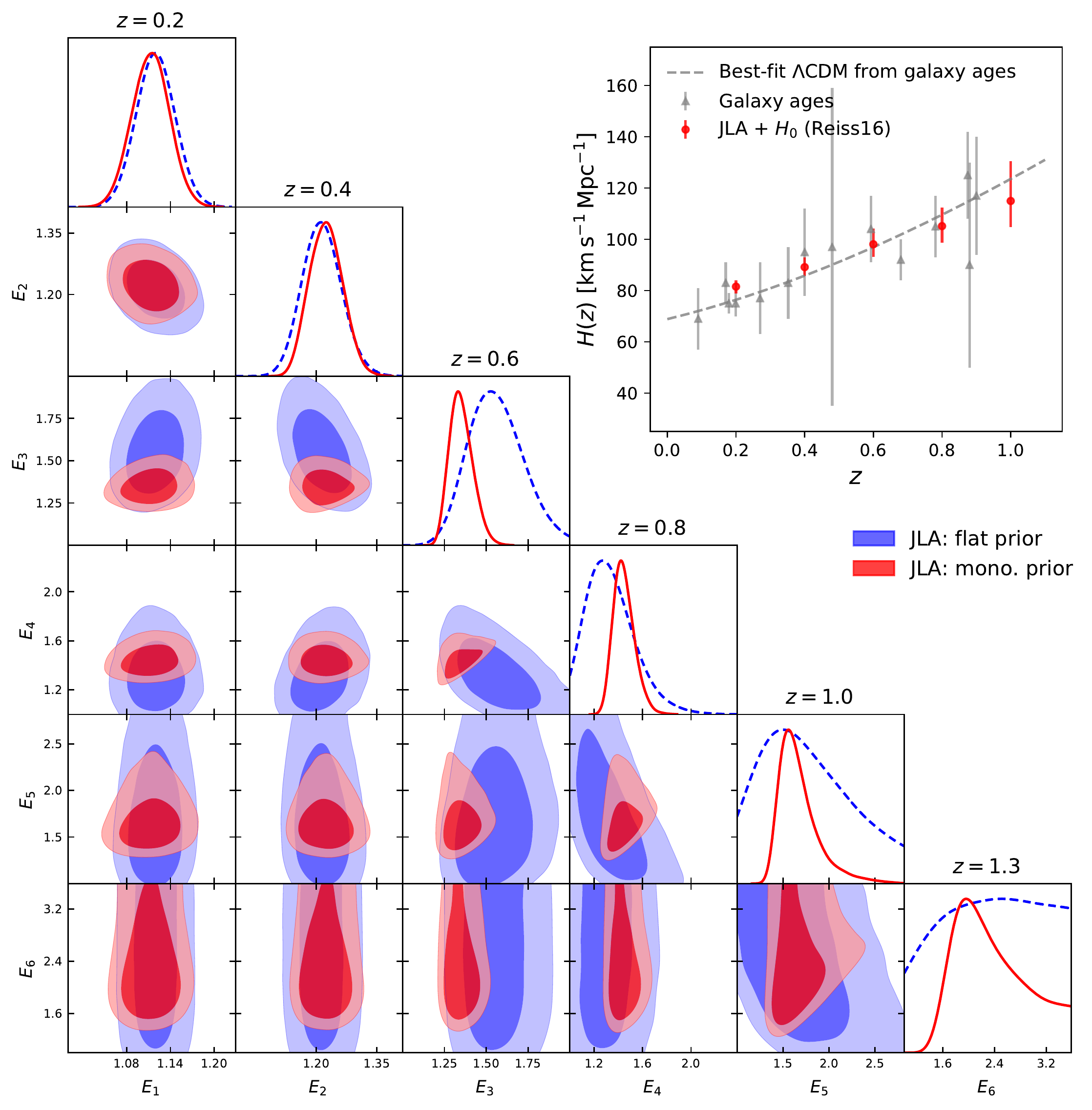}
\caption{Constraints on the cosmic expansion rate $\{E_i\}$ from JLA. The
marginalized one dimensional probability density distributions in the panels
along the diagonal are normalized to the same peak height. The inset shows the
Hubble parameter $\{H_i\}$ (circles), which is mapped from JLA-reconstructed
$\{E_i\}$ and the Reiss16 $H_0$ measurement, along with those derived from
galaxy age data (triangles)~\citep{2002ApJ...573...37J,2010MNRAS.406.2569C}
and the best-fit $\Lambda$CDM $H(z)$ (dashed line) of the galaxy age data.}
\label{cnt_jla}
\end{figure}

Table~\ref{table_USJ_Ei} summarizes the cosmic expansion rate $\{E_i\}$ and the
Hubble parameter $\{H_i\}$ reconstructed from the three SN Ia compilations. The
values of $\{H_i\}$ are mapped from $\{E_i\}$ using the prior
$H_0=73.24 \pm 1.74~\rm{km~s^{-1}~Mpc^{-1}}$, which is based on the combination
of maser distance of NGC 4258, trigonometric parallaxes of Milky Way Cepheids
and Cepheids in the Large Magellanic Cloud~\citep[Riess16]{0004-637X-826-1-56}. Results
at the last interpolation node of each compilation are rather poor (see e.g.,
figure~\ref{cnt_jla}), so they are not  included in table~\ref{table_USJ_Ei}.
The results from Union2.1 and JLA are consistent with each other, whereas the
SNLS3 results at $z=0.4$ and $0.6$ are considerably lower than those from the
other two datasets. Reconciling the differences is beyond the scope of this work. 
Nevertheless, we note that inconsistency also exists between the inferred matter
density $\Omega_{\rm m}$ from SNLS3 and those from Union2.1 and JLA, which has
been attributed to an unanticipated systematic effect \citep{2014A&A...568A..22B}.

Figure~\ref{cnt_jla} presents the cosmic expansion rate $\{E_i\}$ reconstructed
from the JLA sample. The marginalized probability density distributions of
$\{E_i\}$, obtained with and without the monotonicity prior (solid curves and
dashed curves, respectively), are compared in the panels along the diagonal,
showing similar improvements brought by the monotonicity prior as seen in
figure~\ref{fig_mock_jla}. The inset in figure~\ref{cnt_jla} shows the Hubble
parameter $\{H_i\}$ (circles) mapped from $\{E_i\}$ and those derived from the
ages of passively evolving galaxies (triangles) \citep{2002ApJ...573...37J,
2010MNRAS.406.2569C}. The JLA-reconstructed $\{H_i\}$ is consistent with the
predicted $H(z)$ (dashed line) in the best-fit $\Lambda$CDM model of the
galaxy age data, which has best-fit $H_0=68.86~\rm{km~s^{-1}~Mpc^{-1}}$ and
$\Omega_{\rm m}=0.317$.\footnote{The marginalized Hubble constant is
$H_0=68.42\pm 3.28~\rm{km~s^{-1}~Mpc^{-1}}$, being consistent with the
Reiss16 $H_0$ measurement to within $1.3\sigma$.}
Figure~\ref{cnt_jla} and table~\ref{table_USJ_Ei}
demonstrate that, with the help of the monotonicity prior, existing SN Ia samples
already provide fairly well and model independent constraints on the cosmic expansion
history at $z\lesssim 1$.

\subsection{Forecast for WFIRST}
\label{sec_forecast}

Here we give a comparison of the constraints on the expansion history expected
from DESI and WFIRST.  DESI will measure $\{H_i\}$ from the radial BAO signal.
We adopt the results from Table~\uppercase\expandafter{\romannumeral5} of 
ref.~\citep{2014JCAP...05..023F}, which covers $z=0.15$ to 1.85 in intervals of
$\Delta z=0.1$.\footnote{Results from ref.~\citep{2014JCAP...05..023F} are actually
the errors forecasted for $H(z)s$, where $s$ is the co-moving sound horizon at the
decoupling epoch.  Because $s$ can be calibrated to a high precision, much better
than $1\%$, we neglect its contribution to the uncertainties of $E_i$ in eq.~\eqref{err_E}.}
WFIRST has a spectroscopic BAO survey component, but we only use its SN Ia results
on $\{E_i\}$ described in section~\ref{test_method}. 

\begin{figure}[t!]
\centering
\includegraphics[width=0.65\textwidth]{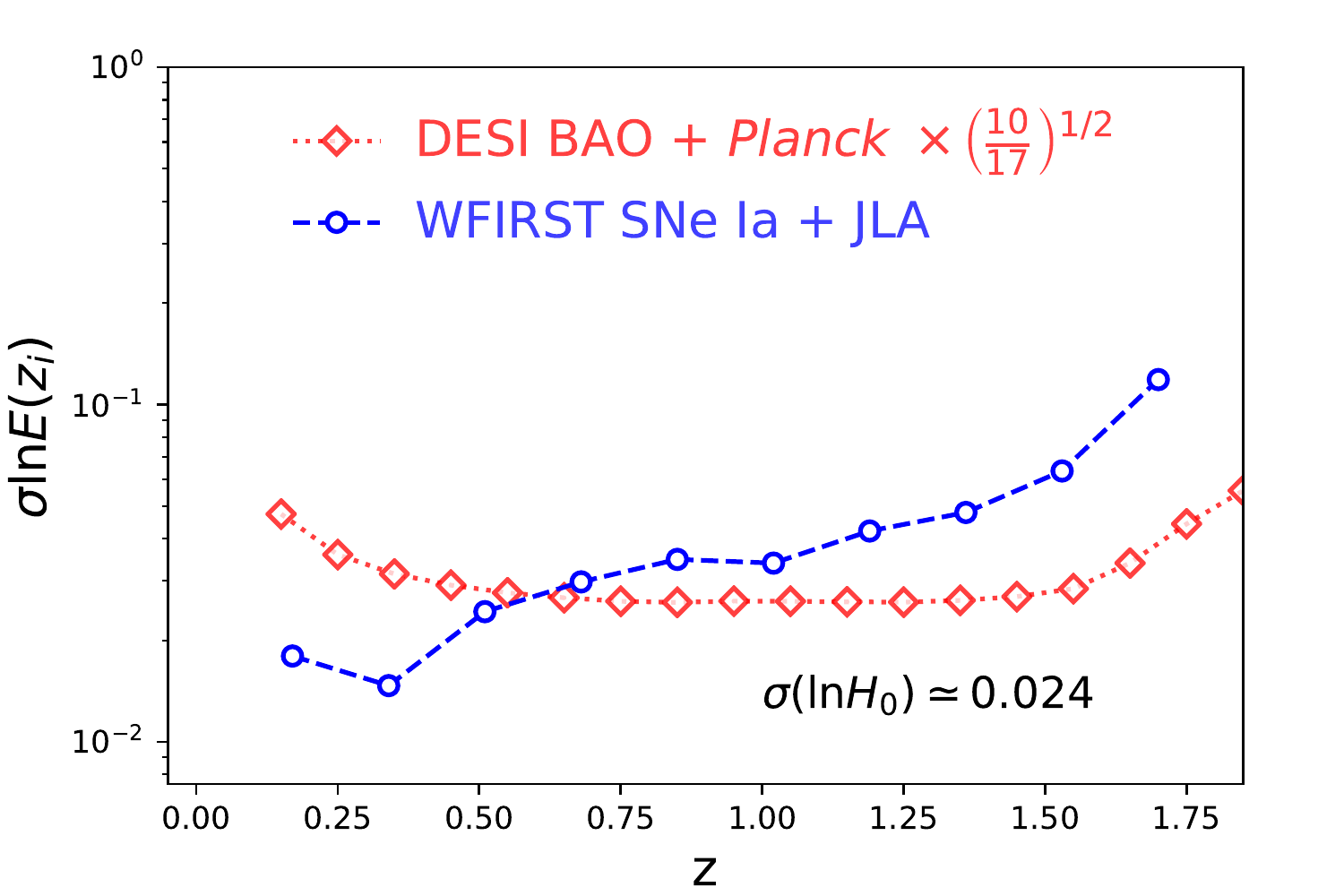}
\caption{Fractional errors of $\{E_i\}$ expected from DESI BAO (diamonds, with
{\it Planck}) and WFIRST SNe Ia (circles, with JLA). The DESI errors have been
rescaled for a proper comparison.}
\label{fig_forecast}
\end{figure}

To compare the results, we convert the uncertainties of $\{H_i\}$ from DESI to
those of $\{E_i\}$ with an independent prior
$\sigma\left(\ln H_0\right) \simeq 0.024$~\citep{0004-637X-826-1-56},
using the following relation
\begin{eqnarray}
\label{err_E}
\sigma^2(\ln E_i) &=& \sigma^2(\ln H_i) + \sigma^2(\ln H_0).
\end{eqnarray}
Moreover, we normalize the errors on $\{\ln H_i\}$ by the square
root of the number of constraints within the same redshift range to compare the
results of different datasets on a roughly equal footing~\citep{Zhan2009}. The multiplicative
factor for DESI is $\left(\frac{10}{17}\right)^{1/2}$, where the numerator in
the parenthesis is the number of interpolation nodes for WFIRST SNe Ia within
$0.1<z\le 1.7$ and the denominator is the number of DESI constraints within roughly
the same redshift range. 

Figure \ref{fig_forecast} shows the expected fractional errors on the expansion
rate from WFIRST SNe Ia and DESI BAO. The DESI results include the CMB prior
from {\it Planck}. The WFIRST-mock results are obtained with the monotonicity
prior enforced, and the JLA-mock is added to strengthen the constraints at the
lowest redshifts.  In terms of $\{E_i\}$, WFIRST SNe Ia can achieve a few percent
level constraints at redshift lower than $z\simeq 1$, fairly competitive to DESI
BAO, though the conclusion depends on the assumed uncertainty of the $H_0$
measurement. At higher redshift, the SN Ia data degrades quickly and become
less competitive than DESI BAO.

\section{Summary}
\label{sec_summary}

In this work we propose a monotonicity prior for non-parametric reconstruction
of the cosmic expansion history, which requires the Hubble parameter $H(z)$ or
the expansion rate $E(z)$ to always increase with redshift. This prior is
reasonably well motivated, even though no theory so far prohibits a reverse of
the trend at some point in the past. Using SN Ia luminosity distances as an
example, we demonstrate that the monotonicity prior is highly effective in
reducing errors on the reconstructed $\{E_i\}$ and that it does not introduce 
appreciable biases. The improvement is most significant at high redshift where
the SN Ia data is poor. It is also helpful that the monotonicity prior keeps
the uncertainties of the reconstructed $\{E_i\}$ fairly independent of the
number of interpolation nodes assigned for the SN Ia data.

Application of the monotonicity prior to existing SN Ia compilations is able
to constrain $\{E_i\}$ to 3\%--6\% at four redshifts up to $z = 0.8$ and 
14\% at $z=1$. When converted to the Hubble parameter using the prior of
$H_0=73.24\pm1.74~\rm{km~s^{-1}~Mpc^{-1}}$, the results are consistent with
those derived from galaxy ages. Future SN Ia sample from WFIRST can achieve
$\lesssim 5\%$ errors on $\{E_i\}$ at redshifts up to $z\sim 1.4$, which
is competitive with DESI BAO at low redshift. Given that WFIRST SNe Ia and
DESI BAO are highly independent of each other, their combination will tightly
constrain $\{E_i\}$ across the entire redshift range of the data. Moreover,
SNe Ia data is sensitive to the expansion rate, while BAO and WL are sensitive
to the Hubble parameter. Hence, the combination of these probes can determine
the Hubble constant as well.

We note that the expansion history reconstructed from SN Ia distances dose
not provide more information than the distances themselves do, but it offers
a convenient way to compare the results of different methods. The monotonicity
prior is helpful for the reconstruction from SN Ia data and should be also
applicable to other probes such as BAOs that measure the Hubble parameter
at multiple redshifts. Even if the expansion history is not directly
reconstructed, one may still enforce the monotonicity prior by requiring
the Hubble parameter or the expansion rate to increase with redshift
monotonically in the MCMC-sampled cosmological models. Better cosmological
constraints could be achieved in this way. Moreover, we expect that the
monotonicity prior can be generalized to other cosmological quantities
that are reasonably monotonic with redshift.

\acknowledgments
This work was supported by the National key Research Program of China
No. 2016YFB1000605 and an internal grant of the National Astronomical 
Observatories of China.

\appendix
\section{An efficient implementation of the monotonicity prior}
\label{prior_eta}

The monotonicity prior on the cosmic expansion history rules out most part
of the parameter space spanned by $\{E_i\}$. However, the MCMC sampler is
unaware of this exclusion zone {\it a priori} and has to test it on the fly.
The MCMC sampling is therefore rather inefficient in $\{E_i\}$ parametrization,
especially when its dimension becomes large. The sampling efficiency can be 
greatly improved by a change of parameters as follows. Let the new set of
parameters $\bm{\eta}\equiv\{\eta_i\}$ be defined through
\begin{equation} \label{eta_E}
E_{i}-1 = \prod_{n=i}^{N}\eta_{n},
\end{equation}
where $N$ is the number of interpolation nodes, $0 < \eta_i \le 1$ 
for $1\le i \le N-1$, and $\eta_N = E_N-1$. Because eq.~\eqref{eta_E}
preserves the relation $E_i \le E_{i+1}$, sampling in $\bm{\eta}$ space
satisfies the monotonicity prior automatically and is thus far more
efficient than that in $\bm{E}$ space. The probability $\pi(\bm{\eta})$
is given by
\begin{equation} \label{pi_eta1}
\pi(\bm{\eta}) 
  = \left| \frac{\partial \bm{E}}{\partial \bm{\eta}} \right| 
 \pi(\bm{E}) = \prod _{i=2}^{N} \eta_{i}^{i-1},
\end{equation}
where we have made use of $\pi(\bm{E})=1$. After a sample of $\bm{\eta}$
is obtained, one can map it into a sample of $\bm{E} $ using 
eq.~\eqref{eta_E}.

\providecommand{\href}[2]{#2}\begingroup\raggedright\endgroup


\end{document}